\documentstyle[aps,twocolumn,epsfig]{revtex}

\begin{document}

\title{Particle--Hole Symmetry and the \\ 
Effect of Disorder on the Mott--Hubbard Insulator}

\author{P.~J.~H. Denteneer} 
\address{Lorentz Institute, Leiden University, P.O. Box 9506, 
2300 RA  Leiden, The Netherlands} 
\author{R.~T. Scalettar} 
\address{Physics Department, University of California, 1 Shields Avenue, 
Davis, CA 95616, USA}
\author{N. Trivedi}
\address{Department of Theoretical Physics, Tata Institute of
Fundamental Research, Homi Bhabha Road, Mumbai 400005, India}

\address{
\begin{minipage}[t]{7.0in}
\begin{abstract}
Recent experiments have emphasized that our understanding of the interplay
of electron correlations and randomness in solids is still incomplete.
We address this important issue and demonstrate that particle-hole (ph) 
symmetry plays a crucial role in determining the effects of disorder on 
the transport and thermodynamic properties of the 
half-filled Hubbard Hamiltonian. 
We show that the low-temperature conductivity decreases with 
increasing disorder when ph-symmetry is preserved, and shows the opposite 
behavior, i.e. conductivity increases with increasing disorder, when 
ph-symmetry is broken.
The Mott insulating gap is insensitive to weak disorder 
when there is ph-symmetry, whereas in its absence the 
gap diminishes with increasing disorder.
\end{abstract}
\end{minipage}}

\maketitle


\vspace*{-1.5cm}

\noindent
{\em Introduction:}
The interplay of disorder and interactions is at the heart
of many interesting and unexplained phenomena in
condensed matter physics. 
For example, the effects of disorder and interactions in 
two-dimensional (2d) electronic systems acting separately lead
to insulating behavior of the Anderson and Mott kinds, respectively.
However, experimental findings on silicon metal-oxide-semiconductor
field-effect transistors (MOSFETs) show the occurrence of a 
conducting phase\cite{AbrahamsRMP}, which is the result of
the combined importance of randomness and interactions.\cite{Fink,PD99}
Other examples of situations in which both disorder and 
interactions are crucial, yet incompletely understood, 
include the formation of local moments 
and the behavior of the susceptibility in doped
semiconductors\cite{BhattLeeSiP}, the superconductor--insulator transition
and universal conductivity in thin metallic films\cite{GoldMar,MPAF}, and
the pinning of flux lines in type--II superconductors.\cite{BlatterRMP}

In recent years, it has become increasingly clear that for
non-interacting electrons the presence or absence of
certain symmetries is crucial in determining the effect of
disorder on both transport and thermodynamic properties, as well
as critical properties of the localization transition.\cite{GadeWegner}
Recent examples where symmetry considerations are important are given 
in the context of quantum wires\cite{BROUWER}
and disordered superconductors,\cite{Senthil,BROUWER2}
where chiral, time--reversal, and spin--rotation symmetries play an
important role.
 
In this paper, we examine the effect of different types of
disorder on both the dynamic and equilibrium thermodynamics of
the 2d Hubbard model in the vicinity of half-filling,
electron density $\langle n \rangle=1$.
We demonstrate that the presence or absence of {\em particle--hole symmetry}
determines the effect of randomness on the conductivity
and the Mott gap.


\vskip0.1in

\noindent
{\em The Model and Computational Approach:}
We consider the following 2d Hubbard Hamiltonian,
\begin{eqnarray}
H &=& - \sum_{\langle {\bf i}{\bf j} \rangle,\sigma } t_{{\bf i}{\bf j}} 
c_{{\bf i}\sigma}^{\dagger} c_{{\bf j}\sigma}^{\phantom \dagger}
- \sum_{\langle\langle {\bf i}{\bf k} \rangle\rangle,\sigma } 
t^{\prime}_{{\bf i}{\bf k}} 
 c_{{\bf i}\sigma}^{\dagger} c_{{\bf k}\sigma}^{\phantom \dagger}
\nonumber \\
&+& U \sum_{{\bf j}} \, (n_{{\bf j} \uparrow}-\frac12) 
(n_{{\bf j} \downarrow}-\frac12)
- \sum_{{\bf j},\sigma} \mu_{\bf j} \, n_{{\bf j} \sigma} ~.
\label{eq:HHub}
\end{eqnarray}
Here $t_{{\bf i}{\bf j}}$ is a bond--dependent hopping matrix element on
near-neighbor sites $\langle {\bf i}{\bf j} \rangle$,
$t^{\prime}_{{\bf i}{\bf k}}$ is a bond--dependent hopping matrix element 
on next-near-neighbor sites 
$\langle\langle {\bf i}{\bf k} \rangle\rangle$,
$U$ is an on--site repulsion,
and $\mu_{\bf j}$ is a site-dependent chemical potential.
We choose
$P(t_{{\bf i}{\bf j}}) = 1/\Delta_t$ for
$t_{{\bf i}{\bf j}} \in [t-\Delta_t/2, t+\Delta_t/2]$, and zero otherwise,
with $t=1$ to set our scale of energy.
Similarly,
$P(t^{\prime}_{{\bf i}{\bf k}}) = 1/\Delta_t^{\prime}$ for
$t^{\prime}_{{\bf i}{\bf k}} \in [t^{\prime} -\Delta_t^{\prime}/2, 
t^{\prime} +\Delta_t^{\prime} /2]$, and
$P(\mu_{{\bf j}}) = 1/\Delta_\mu$ for
$\mu_{{\bf j}} \in [-\Delta_\mu/2, +\Delta_\mu/2]$,
so that the various $\Delta$ measure the disorder strength.
We will focus on half--filling
where the effects of interactions are most prominent,
as evidenced by the formation of antiferromagnetic correlations
and a Mott-Hubbard charge gap at low temperatures.

Our computational technique is determinant Quantum Monte Carlo
(QMC),\cite{WHITE}
an approach which allows us to study much larger numbers of
particles than those explored with exact 
diagonalization (and the two-electron problem).\cite{Vojta,TIP}
In this method the electron--electron interactions are replaced by a 
space and imaginary time dependent Hubbard--Stratonovich field.
The integral over possible field configurations, which exactly
retains the interactions in the problem, is done stochastically
and allows us to calculate static and dynamic (Matsubara time) 
correlation functions at a fixed temperature $T$.

The disordered Hubbard model in Eq.~(\ref{eq:HHub}) 
is particle--hole (ph) symmetric 
when $t^{\prime}_{{\bf i}{\bf k}}=\mu_{\bf j}=0$.  
That is, under the transformation
$c_{{\bf i}\sigma}^{\dagger} \rightarrow (-1)^{\bf i}c_{{\bf i}\sigma}$ 
the Hamiltonian is unchanged, and
the system is precisely half--filled for all values of the parameters
in $H$ and also for all $T$.
Therefore, while near-neighbor bond and local site disorder
both introduce randomness into the system, they differ fundamentally
in that site disorder breaks ph--symmetry.

In order to reveal the effects of disorder quantitatively, we will
examine the transport by evaluating the {\em temperature-dependent} 
dc--conductivity,
\begin{equation}
 \sigma_{\rm dc} \simeq
   \frac{\beta^2}{\pi} \Lambda_{xx} ({\bf q}=0,\tau=\beta/2) ~,
 \label{eq:condform}
\end{equation}
(with $\beta \equiv 1/k_{\rm B}T$) as determined\cite{RANDERIA} from the 
current--current correlation function,
$\Lambda_{xx} ({\bf q},\tau) =
   \langle j_x ({\bf q},\tau) \, j_x (-{\bf q}, 0) \rangle$.
Here $j_x ({\bf q},\tau)$, the ${\bf q},\tau$-dependent current
in the $x$-direction, is the Fourier transform of,
$j_x ({\bf l}) =  i \sum_\sigma \, t_{{\bf l} + \hat{x},{\bf l}}
(c^{\dagger}_{{\bf l} + \hat{x},\sigma}
c^{\phantom \dagger}_{{\bf l}\sigma}
- c^{\dagger}_{{\bf l}\sigma}
c^{\phantom \dagger}_{{\bf l}+\hat{x},\sigma}).$
From the one-electron Green function as a function of
imaginary time we extract the {\em temperature-dependent} density of 
states at the chemical potential $N(\epsilon=0)$\cite{TriRan95}:
\begin{equation}
   N(0) \simeq 
-\beta \, G \left({\bf r} = 0, \, \tau = \beta/2 \right)/\pi  ~.
\label{eq:N0}
\end{equation}
These two quantities allow a clear characterization of the transport
and thermodynamic properties of the system.


\vskip0.1in
\noindent
{\em Results:}
First we discuss the transport properties:
in Fig.~1, we exhibit the effect of near-neighbor hopping 
(bond) disorder on the conductivity.
For all disorder strengths $\Delta_t$, at temperatures greater than 
a characteristic
temperature related to the Mott gap, the system shows metallic behavior
with $\sigma_{\rm dc}$ increasing upon lowering $T$. 
The conductivity turns down sharply as the temperature drops below a
characteristic temperature and the system shows insulating behavior with
$\sigma_{\rm dc}$ decreasing upon lowering $T$.
In the case of zero randomness, the perfect nesting of the Fermi surface
in 2d leads to antiferromagnetic long range order (AFLRO)
in the ground state with  and an associated spin density wave gap
for arbitrarily small $U$ evolving to a Mott gap at larger $U$.
Hopping disorder reduces AFLRO via the formation of singlets
on bonds with large hopping $t_{\bf ij}$ and hence large coupling
$J=t_{\bf ij}^2/U$ and ultimately destroys it beyond 
$\Delta_t \approx 1.6t$.\cite{dishubqmc}
The fascinating result we have found is that insulating behavior
in the conductivity nevertheless persists to much larger
$\Delta_t$. Moreover from the shift of the maximum in Fig.~1
we deduce that the mobility gap in fact increases with increasing
$\Delta_t$.

The situation is quite different in the case of site
disorder, as shown in Fig.~2:
at fixed temperature $T$, as site disorder $\Delta_\mu$ is turned on, 
the conductivity increases, i.e. the Mott insulating state is
weakened.\cite{foot1}
At weak disorder, the conductivity drops with decreasing $T$, 
reflecting again the presence of the Mott insulating phase.
As the disorder strength becomes large enough to neglect
$U$, one would expect a similar temperature dependence arising from
Anderson insulating behavior.
We believe that in all cases the conductivity will ultimately
turn over and go to zero at low $T$, but we are limited in these
simulations to temperatures
$T>W/48$ because of the fermion sign problem.
Nevertheless, the data for site disorder offer a dramatic contrast to that
of bond disorder (Fig.~1) where randomness decreases the conductivity.

\begin{figure}
\begin{center}
\vskip-1.2cm
  \epsfig{figure=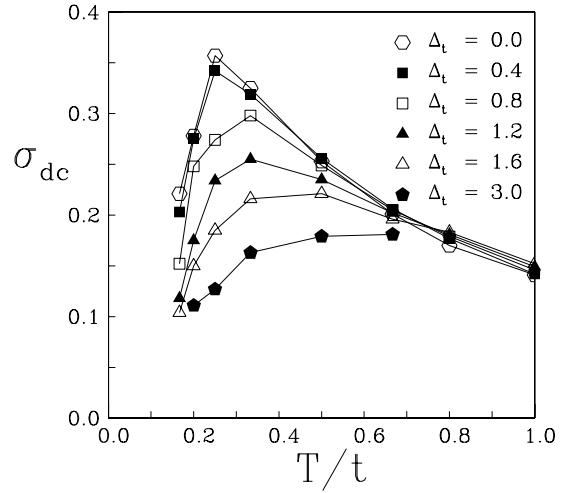,width=\linewidth}
\end{center}
\vskip-0.9cm
 \caption{\label{fig:Fig1} 
The effect of particle--hole--symmetry preserving (near-neighbor) 
bond disorder in the half-filled Hubbard Hamiltonian is to
decrease the conductivity $\sigma_{\rm dc}$. 
Data is for $U=4t$ on a $8\times 8$ square lattice; 
$\Delta_t$ measures the strength of the bond disorder.
}
\end{figure}
\begin{figure}
\begin{center}
\vskip-1.4cm
  \epsfig{figure=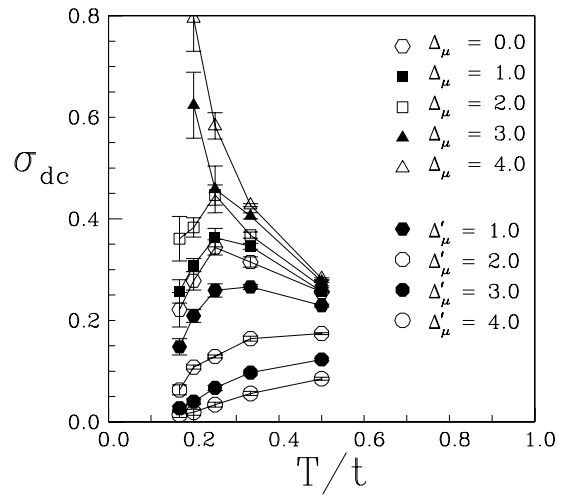,width=\linewidth}
\end{center}
\vskip-0.9cm
 \caption{\label{fig:Fig2} 
Canonical site disorder (with strength $\Delta_\mu$) enhances the 
conductivity. Particle--hole symmetric site disorder (with strength 
$\Delta^{\prime}_\mu$), as with bond disorder (Fig.~1), suppresses 
the conductivity. Other parameters are as in Fig.~1.
}
\end{figure}

What is the underlying reason for the different effects of bond and site 
disorder on the conductivity?
There are several obvious differences in the effect of
bond and site disorder on local and even longer range spin
and charge correlations.
Site disorder enhances the amount of double occupancy on the lattice,
since the energy cost $U$ of double occupancy is compensated by
differences in site energies. One explanation of why site 
disorder increases $\sigma_{dc}$ is that the concomitant increase in 
empty sites leads to more mobility. This destruction of 
local moments ultimately also leads to the end of antiferromagnetic order.
Surprisingly, we find in our simulations that bond disorder
has a similar diminishing effect on local moments, suggesting that the 
difference in the behavior of the conductivity arises from a different
origin.

We argue here that {\em particle--hole symmetry} is the unifying
criterion which underlies and determines the effect of disorder.
As emphasized above, site and bond disorder have
rather similar effects on the double occupancy. Moreover, the consequences
of this effect for $\sigma_{\rm dc}$ are expected to become visible only
above a threshold value of disorder strength, 
whereas we observe effects on $\sigma_{\rm dc}$ already for weak 
disorder. Instead, the key distinction is
in the presence or absence of ph--symmetry.
In order to explore this conjecture more fully, we have studied
two other types of disorder: site disorder that preserves 
ph--symmetry and bond disorder that breaks ph--symmetry 
(by including next-near-neighbor hopping).

{\em Particle--hole symmetric} site disorder is introduced by
adding random chemical potentials to the Hubbard model
which couple with opposite sign to the density of 
up and down electrons, i.e. choose 
$\mu_{\bf j} \equiv \mu_{{\bf j}\sigma} = \sigma\mu_{\bf j}$
in (\ref{eq:HHub}).
This type of disorder represents a random (Zeeman) magnetic field.
For $U=0$ ph--symmetric site disorder has precisely the same effect 
as conventional site disorder, since moving in a given random chemical 
potential landscape or one obtained by reversing all the site
energies is entirely equivalent.
However the behavior of the conductivity at finite $U$ is 
dramatically different. 
Fig.~2 shows that ph--symmetric site disorder (with strength 
$\Delta^{\prime}_\mu$) has the {\em same} effect on $\sigma_{\rm dc}$
as bond disorder, i.e. conductivity decreases with increasing  
$\Delta^{\prime}_\mu$.

To seek final confirmation of our conjecture, 
we have also explored the effect of next-near-neighbor (nnn) hopping 
and randomness therein.  
Such longer ranging hybridization breaks ph--symmetry
on a square lattice, since it connects sites on the same sublattice.
We find that such disorder has the {\em same} effect as conventional 
site randomness, i.e. increases the conductivity at finite $T$. 
Thus in all four types of disorder, the behavior of the
conductivity falls into the appropriate class based on the
preservation or destruction of ph--symmetry,
strengthening the evidence that it is this symmetry
which is playing the crucial role in determining the effect
of randomness on the transport properties.

We now turn to thermodynamic properties.
The most direct measure of the Mott gap is from the compressibility, 
or from the behavior of density
$\langle n \rangle$ as a function of chemical potential $\mu$, as
shown in Fig.~3. The range of $\mu$ where $\langle n \rangle$ is
constant (and the system is incompressible) is a direct measure
of the gap in the spectrum.
Hopping and ph-symmetric site disorder clearly stabilize the
plateau of the density at half--filling.
On the other hand, conventional site disorder (with $\Delta_\mu=U/2$)
has a compressibility which is indistinguishable 
(within the computational possibilities) from the clean system.

\vskip-0.5cm
\begin{figure}
\begin{center}
  \epsfig{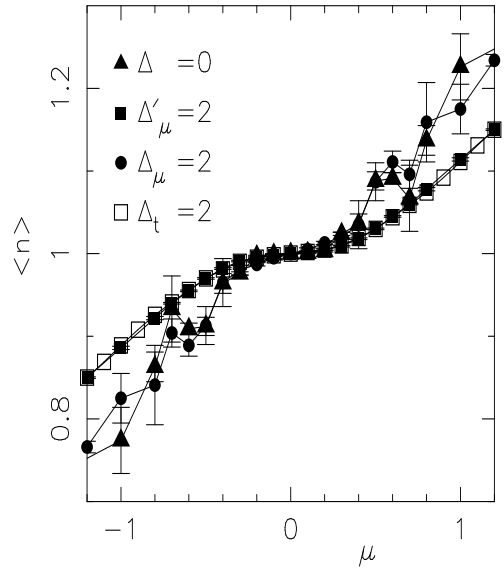}
\end{center}
\vskip-0.1cm
 \caption{\label{fig:Fig3} 
The Mott gap is made more robust by the addition of bond disorder or
particle-hole symmetric site disorder (open and filled squares)
of strength $\Delta=2t=U/2$,
as indicated by the response of the 
density to changes in the chemical potential.
For canonical site disorder (filled circle) the Mott gap
is practically unaffected by this strength of randomness. [19]
Calculations are for $T = t/8 = W/64$ on a $8\times 8$ lattice.
}
\end{figure}

\begin{figure}
\begin{center}
\vskip-1.3cm
\epsfig{figure=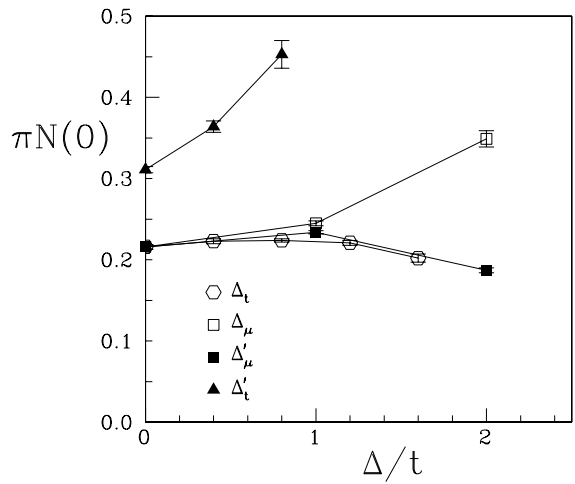,width=\linewidth}
\end{center}
\vskip-0.9cm
\caption{\label{fig:dos}
Behavior of the density of states $\pi N(0)$
at the Fermi level and at fixed low temperature as a function of 
disorder strength $\Delta/t$ for various types of disorder.
All data are for $T=t/6$, except
data for randomness in next-near-neighbor hopping 
(disorder strength $\Delta^{\prime}_t$) which are at 
temperature $T=t/5$ (the value $t^\prime = 0$ is used). [19]
Other parameters are as in Figs.~1 and 2.
}
\end{figure}

The density of states (DOS) at the Fermi level $N(0)$ gives valuable 
information  on the effect of disorder on the Mott gap.
In the pure system, QMC studies have shown that the DOS exhibits
a clear Mott gap with $N(0) \rightarrow 0$ as $T$ is lowered to zero.
The nonzero values of $N(0)$ we obtain at nonzero $T$ reflect the small
residual slopes in the plateaus in the $\langle n \rangle$ vs. $\mu$
plot (cf. Fig.~3); at lower $T$, $N(0)$ approaches zero just as the
plateaus become perfectly flat.
The behavior of $N(0)$ at a fixed low $T$ as a function of the strength 
of the various types of disorder is given in Fig.~4. 
$N(0)$ is rather insensitive to ph--symmetric disorder ($\Delta_t$ and 
$\Delta^{\prime}_\mu$) and is even reduced by it: the Mott gap persists.
On the other hand, ph--symmetry breaking disorder ($\Delta_\mu$ and 
$\Delta^{\prime}_t$) clearly enhances $N(0)$, i.e. fills up the Mott gap.

Our results provide a clear numerical demonstration of the key role
of particle--hole symmetry. The effects can also be understood
qualitatively as follows:
In the clean case, at $\langle n \rangle =1$ and strong coupling, 
the DOS consists of an occupied lower Hubbard band (LHB) and an 
unoccupied upper Hubbard band (UHB),
separated by a charge gap of the order of $U$.
In the case of ph--symmetric disorder, the effect of disorder on
LHB and UHB is identical. Therefore the Fermi energy remains in the 
middle of the gap: this enables the insulating 
behavior and Mott gap to stay intact.
A stabilized charge gap for ph--symmetric {\em site} disorder is evident
since double occupation is strongly suppressed. For nn-hopping disorder
a simple argument is less obvious, but the data in Fig.~3 clearly
show that these two cases fall into the same class.
When ph-symmetry is broken, the LHB and the UHB
will be affected differently; different numbers of states will
appear at either side of the gap. As a consequence, the Fermi energy 
ends up in one of the tails of the DOS, resulting in 
an enhanced $N(0)$ (cf. Fig.~4) and increased conductivity (Fig.~2).
The fact that the states introduced by disorder are 
localized \cite{OTSUKA} will keep the system in an 
insulating state (cf. Fig.~2).

 
\vskip0.1in
\noindent
{\em Conclusions:}
In this Letter, we have shown that particle--hole (ph) symmetry
plays a decisive role in determining the effect of randomness
on transport and thermodynamic properties of the half-filled
Hubbard model.
By exploring four different types of disorder, of which two preserve
ph symmetry and two break it, we demonstrate
that a classification by this symmetry allows us to
understand the effect of disorder on $\sigma_{\rm dc}(T)$,
the charge gap and the compressibility.
The presence of ph symmetry is found to have a {\em protective}
influence on the charge gap.

A related example where symmetry plays a crucial role in the
effects of disorder is the case of 
localization in the superconducting phase, where the
quasiparticles are described by a Bogoliubov--de
Gennes Hamiltonian.\cite{BROUWER2}
In this case, one can classify the system according to the
presence or absence of time reversal and spin rotation symmetries,
and it is found in one dimension
that in the absence of spin rotation symmetry,
the conductance decays algebraically with system size, while in the 
symmetric case it decays exponentially.
Therefore, in this situation as well, the extra spin rotation symmetry
leads to a strengthening of insulating behavior.

The question of the behavior of the half--filled fermion Hubbard model
as disorder is added is furthermore reminiscent of similar
issues in the ph-symmetric boson Hubbard model.\cite{MPAF}
At generic densities,
it is believed that a new 'Bose glass' phase
arises to intervene in the original ground state phase diagram between
superfluid and Mott insulating phases,
but the situation at the ph-symmetric tip
of the Mott lobe is uniquely different.
Our work is a first step in the analysis of the nature of the
behavior of the fermionic model.

The authors thank Piet Brouwer, Andreas Ludwig,
and George Sawatzky for helpful discussions. The research of RTS is
supported by grant NSF--DMR--9985978


\begin{references}
\bibitem{foot1}
We have verified that beyond a certain large site disorder 
($\Delta_\mu \geq 4t$) $\sigma_{\rm dc}$ decreases again.
\bibitem{foot2}
Particle--hole symmetric disorder has no sign problems at half--filling;
data are better then and simulations can go to lower $T$.
\end{references}
\end{document}